\documentclass[12pt,preprint,epsf]{aastex}

\newcommand{\jb}{Jy~beam$^{-1}$}
\newcommand{\kms}{km~s$^{-1}$}
\newcommand{\et}{et~al.}
\newcommand{\muG}{$\mu$G}
\newcommand{\evcm}{eV~cm$^{-3}$}
\newcommand{\ergs}{ergs~s$^{-1}$}

\shortauthors{LaRosa \et\/}
\shorttitle{Weak Magnetic Field Galactic Center Region}

\begin{document}

\title{Evidence of a Weak Galactic Center Magnetic Field from 
  Diffuse Low Frequency Nonthermal Radio Emission}

\author{T.~N.~LaRosa\altaffilmark{1}, C.~L. Brogan\altaffilmark{2},
S.~N.~Shore\altaffilmark{3}, T.~J.~W.~Lazio\altaffilmark{4},
N.~E.~Kassim\altaffilmark{4}, \& M.~E.~Nord\altaffilmark{4,5}}

\altaffiltext{1}{Department of Biological \& Physical
Sciences, Kennesaw State University, 1000 Chastain Rd., Kennesaw, GA
30144; ted@avatar.kennesaw.edu}

\altaffiltext{2}{Institute for Astronomy, 640 North A`ohoku Place, Hilo, HI 96720; cbrogan@ifa.hawaii.edu.}

\altaffiltext{3}{Dipartimento di Fisica ``Enrico Fermi'', Universit\`a di Pisa
  and INFN, Sezione di Pisa, Largo B. Pontecorvo 3, I-56127 Pisa, 
Italy; shore@df.unipi.it}

\altaffiltext{4}{Remote Sensing Division, Naval Research Laboratory,
Washington DC 20375-5351; Joseph.Lazio@nrl.navy.mil; 
Namir.Kassim@nrl.navy.mil; Michael.Nord@nrl.navy.mil.}

\altaffiltext{5}{Department of Physics and Astronomy,
University of New Mexico, Albuquerque, NM 87131.}

\begin{abstract}

New low-frequency 74 and 330 MHz observations of the Galactic center
(GC) region reveal the presence of a large-scale ($6\arcdeg\times
2\arcdeg$) diffuse source of nonthermal synchrotron emission.  A
minimum energy analysis of this emission yields a total energy of
$\sim (\phi^{4/7}f^{3/7})\times 10^{52}$ ergs and a magnetic field
strength of $\sim 6(\phi/f)^{2/7}$ \muG\/ (where $\phi$ is the proton to
electron energy ratio and $f$ is the filling factor of the synchrotron
emitting gas).  The equipartition particle energy density is
$1.2(\phi/f)^{2/7}$ \evcm\/, a value consistent with cosmic-ray data.
However, the derived magnetic field is several orders of magnitude
below the 1 mG field commonly invoked for the GC.  With this field the source can be maintained with the SN rate inferred from the GC star formation.  Furthermore, a strong magnetic field implies an abnormally low GC
cosmic-ray energy density.  We conclude that the mean magnetic field 
in the GC region must be weak, of order 10 \muG\/ (at least on 
size scales $\ga 125\arcsec$). 

\end{abstract}


\section{INTRODUCTION}

An outstanding question in Galactic center (GC) studies concerns the origin,
strength, and role of magnetic fields in the region.  Large-scale
filamentary non-thermal structures (two dozen confirmed), with lengths
of up to tens of parsecs, have been observed in the vicinity of the GC
for over the last two decades (e.g. Morris \& Serabyn 1996, and
references therein; Lang et al. 1999; LaRosa, Lazio, \& Kassim 2001;
Nord \et\/ 2004; LaRosa \et\/ 2004; Yusef-Zadeh, Hewitt, \& Cotton
2004).  The spatial distribution of the non-thermal filaments (NTFs)
is confined to within $\sim 1.5\arcdeg$ of the \hbox{GC}, and this phenomenon
seems to be unique to the GC region. The NTFs are widely believed to
be magnetic field lines illuminated by the injection of relativistic
particles. Within the context of this picture, it has been suggested
that a pervasive, strong ($\sim 1$ mG) magnetic field must permeate
the entire GC region in order to confine the NTFs (e.g., Morris \&
Serabyn~1996 and references therein).  This Letter reports the
discovery of a previously unrecognized diffuse, nonthermal structure
in the GC and shows that its properties are not consistent with a
strong, space-filling magnetic field.

\section{OBSERVATIONS AND RESULTS}

We have used the Very Large Array\footnote{%
The Very Large Array and the Green Bank Telescope are facilities of
the National Radio Astronomy Observatory operated under a cooperative
agreement with the National Science Foundation.}
(VLA) in all four configurations to image the Galactic center region
at 74 MHz.  The resulting image presented in Figure 1\textit{a} has a
resolution of $125\arcsec$, an rms noise of $\sim 0.2$ \jb\/, and a
dynamic range of $\sim 200$.  The details of the 74 MHz data reduction
are discussed more thoroughly in Brogan \et\/ (2003; C.~L.~Brogan et
al.~2005, in preparation).  This is the highest resolution and
sensitivity image of the Galactic center region at frequencies below
300 MHz yet created. Even so, the data used to make Figure 1\textit{a} were
tapered to provide surface brightness sensitivity optimized to the
large-scale emission which is the focus of this Letter.  Future
improvements in ionospheric calibration offer the opportunity to
achieve the full resolving power of the VLA at 74 MHz and to produce
a map more suited to studying smaller, discrete sources.

A wide range of discrete emission and thermal absorption features are
evident in the 74 MHz image (Brogan \et\/ 2003; 2005), as well as a
large-scale region of diffuse emission surrounding Sgr~A and extending
$\pm 3\arcdeg$ in longitude and $\pm 1\arcdeg$ in latitude.  At a
distance of~8~kpc this angular scale corresponds to a region 840 by
280 pc.  Interestingly, none of the NTFs are detected in the 74 MHz
image, and away from known supernova remnants, the emission is quite
smooth. The extent of the large-scale low-frequency structure is well
matched in size to that of the central molecular zone (CMZ); the
region surrounding the Galactic center where the molecular gas
density, temperature, and velocity dispersion are high (Morris \&
Serabyn 1996; Brogan et al. 2003).  We hereafter refer to the
low`frequency structure as the Galactic center diffuse nonthermal
source (DNS).  The even larger scale, smooth Galactic background is
resolved out of this image due to the spatial filtering afforded by
the interferometer; only structures smaller than $\sim 5.5\arcdeg$
(the angular scale corresponding to the shortest baseline) down to the
resolution limit of $125\arcsec$ are sampled.  Although the size of
the DNS is comparable to the largest angular scale to which our
measurements are sensitive, two of its characteristics strengthen our
confidence that it is a distinct source: (1) the strong thermal
absorption in the vicinity of the GC itself (Fig.~1\textit{a})
effectively segregates the DNS into two pieces, each of which is
considerably smaller than $5.5\arcdeg$, and (2) the DNS is also
detected in higher frequency (330 MHz) single-dish data (see below).

To complement our 74 MHz data, we have imaged a $15\arcdeg\times
15\arcdeg$ region centered on the GC with the Green Bank Telescope
(GBT) at 330 MHz.  The GBT data were obtained with a 20 MHz bandwidth,
1024 spectral channels, and a scanning rate that more than Nyquist
sampled the beam ($38.9\arcmin$). The GC region is bright enough at
330 MHz that attenuators are required to avoid saturating the
detectors.  The effect of this high attenuation setting was carefully
calibrated out of the flux calibration scans on 3C~286 and the
observations of the off-positions. In order to correct for the non
zero temperature of the off-positions located at $(l,b)=(\pm
20\arcdeg,0\arcdeg)$, estimates for the sky brightness at these two
locations were estimated from the 408 MHz survey image (Haslam \et\/
1982) assuming a spectral index of $-2.7$ (Platania et al.\ 2003) and
added back to the 330 MHz image. Finally, the image was converted to a
flux density scale using the observed antenna temperature of 3C~286
compared to its expected brightness temperature (Ott \et\/ 1994) and
the gain of the GBT (approximately 2 K/Jy at low frequencies).

A sub image of the resulting GBT 330 MHz image with a resolution of
$38.9\arcmin$ is shown in Figure~1\textit{b}. This image includes
contributions from all of the discrete sources apparent in the
high-resolution VLA 330 MHz image by LaRosa \et\ (2000), the DNS, and
the smooth Galactic background synchrotron emission, with the latter
contributing much of the total flux. Thus, the Galactic background
must be removed and the discrete source contribution accounted for
before the 330 MHz properties of the DNS can be assessed.  For
background subtraction we chose a constant longitude slice free of
discrete sources (and well outside of the DNS) near
$\ell=354.5\arcdeg$ and subtracted it from every other constant
longitude plane (median weight filtering was not used since it can
introduce structures on size scales equal to the filter).  The image
resulting from this procedure is shown in Figure~1\textit{c}. Despite
the resolution difference, there is excellent agreement between the
diffuse structure visible in the background-subtracted GBT 330 MHz
image (Fig.~1\textit{c}) and the 74 MHz VLA image (Fig.~1\textit{a}).
  
Having established the reality of the DNS, we now estimate its
integrated 74 and 330 MHz flux and spectral index.  Due to the copious
thermal absorption at 74 MHz apparent in Fig.~1\textit{a}, a simple
integration of the total flux within the DNS region is
impossible. Instead, we have used the average 74 MHz flux density at
locations within the DNS that appear free of both discrete emission
sources and thermal absorption, together with its apparent size (an
ellipse of dimension $6\arcdeg\times 2\arcdeg$) to calculate its total
flux.  Using this method we find an integrated 74 MHz flux density for
the DNS of $16.2\pm 1$ kJy. This estimate is likely to be a lower
limit, however, as we show below an underestimate of the total 74 MHz
flux density of the DNS does not significantly affect our arguments
regarding the weakness of the large-scale GC magnetic field.

The integrated 330 MHz flux density within the boundaries of the DNS
(Fig.~1\textit{c}) is $\sim 8000$ Jy, while the total flux density
contained in discrete sources from the VLA 330 MHz image (LaRosa
\et~2000) is $\sim 1000$ Jy.\footnote{%
The VLA 330 MHz image is not sensitive to structures larger than $\sim
1\arcdeg$ and thus contains little or no contribution from the
Galactic background or the DNS.}
Thus, we estimate that the integrated 330 MHz flux density of the DNS
is $\sim 7000$ Jy. Except for the immediate vicinity of the Galactic
center itself (where the {\em thermal} ionized gas density is very
high), thermal absorption effects should not be a significant effect
at 330 MHz (e.g., Nord \et\ 2004) Combined with the lower
limit to the 74 MHz DNS flux, we find that the spectral index of the
DNS must be steeper than $-0.7$ (assuming
$S_{\nu}\propto\nu^{\alpha}$) or $-2.7$ if brightness temperatures are
considered.  This value is comparable to the brightness temperature
spectral indices measured for the Galactic plane synchrotron
background emission ($-2.55$ to $-2.7$; see Platania \et\ 2003).

\subsection{Minimum Energy Analysis}

Using standard synchrotron theory (Moffat 1975), the minimum energy
and magnetic field are given by $U_{min}=0.5(\phi AL)^{4/7}V^{3/7}$
and $B(U_{min})=2.3(\phi AL/V)^{2/7}$.  Here $A$ is a function of the
spectral index, $L$ is the luminosity, $V$ is the source volume, and
$\phi$ is the ratio of energy in protons compared to electrons.  Using
a 74/330 MHz spectral index of $-0.7$ and integrating from 10 MHz to
100 GHz we find a total luminosity for the DNS of $3\times 10^{36}$
\ergs\/ (we have used $S_{330}=7000$ Jy and
$S_{\nu}=S_{330}\nu^{-0.7}$ for this calculation).  If, for example,
the spectral index were actually as steep as $-1.5$ the luminosity
would still be $> 10^{36}$ \ergs\/. Thus, the possibility that we have
underestimated the 74 MHz flux is unlikely to significantly impact
this analysis. Assuming that the depth of the DNS is 480 pc (the
geometric mean diameter assuming a distance of 8 kpc), the minimum
energy is $(\phi^{4/7}f^{3/7})\times 10^{52}$ ergs and the
corresponding magnetic field strength is $6(\phi/f)^{2/7}$ \muG\/,
where $f$ is the filling factor of the nonthermal emission within the
volume $V$.  With these values, the equipartition particle energy
density is $1.2(\phi/f)^{2/7}$ \evcm\/.  A filling factor as low as
1\% will only increase the derived field strength and particle density
by a factor of $\sim 4$.  A large proton to electron energy ratio of
100 could increase these estimates by another factor of $\sim 4$.
Thus, even with these extreme parameters, the magnetic field on size
scales larger than the 74 MHz beam ($125\arcsec$) must be $\la 100$
\muG\/.

The above calculation applies to the entire $6\arcdeg\times 2\arcdeg$
region spanned by the DNS.  However, the nonthermal filaments are
found only in the inner $1.5\arcdeg\times 0.5\arcdeg$.  The integrated
330 MHz flux in this smaller region is $\sim 1000$ Jy and yields a
minimum energy of $(\phi^{4/7}f^{3/7})\times 10^{51}$ ergs (electron
energy density of $\sim 7.2$~eV~cm${}^{^3}$) and magnetic field of
$11(\phi/f)^{2/7}$ \muG\/.  Thus, the gradient in the DNS brightness
does not translate into a significantly larger magnetic field in the
innermost region compared to the region as a whole.

\section{DISCUSSION}

\subsection{Energy Requirements}

For the minimum energy parameters the radiative lifetime of electrons
generating the DNS is of order $5 \times 10^7$~yr while electrons in a
strong 1 mG magnetic field would have a much shorter synchrotron
lifetime, only $10^5$~yr.  The energy contained in a typical supernova
remnant in particles and magnetic field is $5 \times 10^{49}$ ergs
(Duric et al 1995), so that $\sim 200$ supernovae (SNe) within the
last $5 \times 10^7$~yr are required to power the DNS.  The SN rate
scales with the star formation rate, which is a factor of $\approx
250$ higher in the inner 50 pc than in the disk (Figer et al 2004).
Given this enhanced star formation rate, the GC SN rate of about {\em
one} SN per $10^5$~yr is sufficient to power the DNS and maintain it
in equilibrium.  Alternatively the DNS could be due to a single
extreme event (with an energy of $\sim 10^{52}$ ergs) or was formed
during a \emph{recent} starburst.  Indeed, the presence of a large
bipolar wind structure centered on the GC, as detected in dust and
radio emission studies, suggests that the GC has undergone periods of
more prolific star formation in the past (e.g., Sofue \& Handa 1984;
Bland-Hawthorn \& Cohen 2003).  However, Bland-Hawthorn \&
Cohen~(2003) estimate that the GC ``Omega lobe'' was formed during a
starburst event $10^7$~yr ago, inconsistent with the 1 mG synchrotron
lifetime.  Additionally, the agreement between the derived DNS
spectral index ($-0.7$) and that of the diffuse Galactic plane
emission suggests that the acceleration mechanism is similar.  This
steep spectral index also suggests that the emission arises from a
relatively old population of electrons.

This scenario for the generation of the DNS is also consistent with
the diffuse soft X-ray emission from the GC region.  Muno \et\/ (2004)
report the presence of diffuse hard (kT$\sim 8$ keV) and soft (kT$\sim
0.8$ keV) X-ray emission from the inner 20 pc. The soft component
requires input of the energy equivalent of one SN every $10^5$~yr, in
good agreement with the rate required to explain the DNS.  The hard
component is more difficult to explain with this injection rate
(requiring $\sim 1$ SN per 3000~yr), but Muno et al. suggest that a
significant fraction of the hard emission may arise from unresolved
cataclysmic variables and other compact objects, so that the high
implied rate may be overestimated.

\subsection{Constraints on GC Cosmic Ray Energy Density}

By construction, the minimum energy procedure constrains the field and
particle energies together but they can also be analyzed separately.
Consider the inner $1.5\arcdeg\times 0.5\arcdeg$ were the magnetic
field if often assumed to be 1 mG.  The integrated flux of the DNS in
this region is $\sim 1000$ Jy.  Assuming a 1 mG magnetic field, this
flux requires a cosmic-ray (CR) electron energy density of 0.04
\evcm\/.  In contrast, the energy density of electrons in the local
interstellar medium (ISM) is a factor of 5 larger (0.2~\evcm; Webber
1998).  Thus, unless the electron energy density in the GC region is
significantly lower than that in other parts of the Galaxy, a 1 mG
field is inconsistent with the 330 MHz data.

The cosmic-ray energy density in a particular region is determined by
the local balance between CR production and escape rates. In the disk,
particles escape with a typical turbulent diffusion velocity of order
10--15 \kms\/ (e.g., Wentzel 1974).  In contrast, it is likely that
particles in the GC region escape much more rapidly due to strong
winds of order a few thousand kilometers per second (e.g., Suchkov, Allen, \&
Heckman 1993; Koyama \et\/ 1996).  However, since the production rate
scales with the SN rate, which exceeds that of the disk by a factor
of $\sim 250$ in the GC region, the higher escape rate in the GC could
be balanced by the higher production rate.  Thus, the cosmic-ray energy
density in the Galactic center region may well be similar to that
measured in the disk.  Unfortunately, no direct {\em measurements} of
the GC cosmic-ray energy density currently exist. There are, however,
a number of indirect indicators that can be used to constrain its
properties.

Diffuse $\gamma$-ray emission in our Galaxy is produced by the
interaction of high-energy cosmic rays with interstellar material, as
well as a relatively uncertain contribution from unresolved compact
sources (e.g. pulsars, X-ray binaries, etc.). Thus, $\gamma$-ray
emission can be expected to peak where the cosmic-ray density is high,
the gas density is high, or there is a high concentration of compact
objects. Recent reanalysis of all available EGRET data toward the GC
region by Mayer-Hasslewander \et\/ (1998) suggest a number of
pertinent conclusions: (1) the level of $\gamma$-ray emission within
the DNS region is entirely consistent with that predicted by models of
the Galactic disk (Hunter \et\/ 1997) with the exception of a $\la
0.6\arcdeg$ radius region of enhanced emission centered on the GC
itself. (2) Taking into account newer estimates of the GC gas mass,
this finding disputes earlier claims that there is a {\em deficit} of
cosmic rays toward the GC region (e.g. Blitz 1985). (3) No excess is
observed toward the Sgr~\hbox{B}, Sgr~\hbox{C}, or Sgr~D molecular cloud complexes as
might be expected if gas density played a large role in the GC
$\gamma$-ray emissivity.  (4) The spectrum of the {\em excess} GC
$\gamma$-ray emission is significantly harder than that of the
Galactic disk and most likely arises from the GC radio arc and/or
compact objects like pulsars.

Galactic center molecular clouds have significantly higher
temperatures, densities, and turbulent velocities than their disk
counterparts (i.e. the CMZ; Morris \& Serabyn 1996;
Rodr{\'\i}guez-Fern{\' a}ndez \et\/ 2004). In the past, heating by a
larger than average CR energy flux has been invoked to explain the
high temperatures of the GC clouds (Suchkov \et\/ 1993). More
recently, a number of studies have shown that the required heating can
be produced by turbulent dissipation, shear due to the intense GC
gravitational potential, and large-scale shocks (e.g.,
Rodr{\'\i}guez-Fern{\' a}ndez \et\/ 2004; G\"usten, \& Philipp 2004).
Moreover, recent observations of the GC region in infrared transitions
of H$_3^+$ by Goto \et\/ (2002) suggest that the GC CR ionization rate
($\zeta$) is consistent with that of the disk $\zeta\sim 3\times
10^{-17}$ s$^{-1}$cm$^{-3}$. It has also been suggested that an
enhanced GC CR density might reveal itself through enhanced abundances
of lithium and boron. These atoms can be formed through spallation
reactions between cosmic rays and the ISM.  Lubowich, Turner, \& Hobbs
(1998) have searched the GC region for hyperfine-structure lines of
neutral Li and B with no success.  From their detection upper limits,
these authors suggest that the GC cosmic ray energy density cannot
exceed the disk value by more than a factor of 13.

Together these results ranging from $\gamma$-ray emission to the
non-detection of lithium and boron provide strong circumstantial
evidence that the GC cosmic-ray energy density is similar to that of
the Galactic disk.

\section{CONCLUSIONS}

Utilizing new 74 and 330 MHz observations, we have discovered a new
Galactic center feature, the diffuse nonthermal source.  The DNS
is extended along the Galactic plane with a size of $\sim
6\arcdeg\times 2\arcdeg$ and a spectral index steeper than $-0.7$. A
minimum energy analysis of the DNS requires that any \emph{pervasive}
magnetic field must be weak, of order 10 \muG\/ (and almost certainly
$\la 100$ \muG\/).  This field strength is 2 orders of magnitude
less than the commonly cited value of 1 mG inferred from the
assumption that the NTFs are tracing a globally organized magnetic
field.  We find that the minimum energy required to fuel the DNS can
be supplied by the enhanced star formation, and hence supernova,
rate within the Galactic center region.

The low global GC region magnetic field derived from this work is
supported by a number of other lines of evidence.  If the global GC 
field strength is 1~\hbox{mG}, then the
observed emission would imply a very low GC CR energy
density. However, EGRET $\gamma$-ray observations (Mayer-Hasslewander
\et\/ 1998; Hunter \et\/ 1997) indicate a GC CR energy density
consistent with that measured in the Galactic disk. Light element (Li
and B) abundance upper limits, $H_3^+$ detections, and consideration
of molecular cloud heating are also all consistent with a normal
Galactic plane cosmic-ray energy density in the GC.

This weak field picture is substantially different from the canonical
one that has emerged over the past two decades for the ``Galactic
center magnetosphere'' (e.g. Morris \& Serabyn 1996; G\"usten, \&
Philipp 2004) wherein the entire Galactic center
region (approximately bounded by the CMZ) has a globally organized
strong magnetic field of order 1 mG.  Together, the radio and
$\gamma$-ray data yield a GC cosmic-ray energy density and magnetic field
strength that are comparable to their
disk values.  Any departures in the energetics from the disk values result from the enhanced level of star formation and associated activity at the Galactic center, an even more extreme version of which may be the nuclear starburst in M82 (Odegard and Seaquist 1991).  It is important to note that our analysis
constrains the average magnetic field strength on size scales larger
than our 74 MHz resolution of $125\arcsec$ and {\em does not} preclude
locally strong magnetic fields on smaller size scales.
         
\acknowledgments
We thank Doug Finkbeiner for an especially useful comment from a critical reading of the manuscript. Basic research in radio astronomy at the NRL is supported by the
Office of Naval Research.  SNS thanks P. Caselli and D. Galli for
discussions.  TNL thanks the INFN/Pisa for travel support.

\begin{figure}[h!] 
\centering 
\caption{
(\textit{a}) VLA image of the Galactic center at 74 MHz with
$125\arcsec$ resolution.
(\textit{b}) GBT image of the Galactic center at 330 MHz with
$40\arcmin$ resolution (beam is shown in lower right corner).
(\textit{c}) Background subtracted GBT image of the Galactic center
region (beam is shown in lower right corner). Some of the brighter emission and free-free absorption sources are indicated in (\textit{a}).}
\end{figure} 

\end{document}